\documentclass[]{interact}

\usepackage{epstopdf}% To incorporate .eps illustrations using PDFLaTeX, etc.
\usepackage[caption=false]{subfig}% Support for small, `sub' figures and tables
\usepackage[natbibapa,nodoi]{apacite}
\theoremstyle{plain}% Theorem-like structures provided by amsthm.sty

\theoremstyle{definition}

\theoremstyle{remark}

\begin{document}
	
	% \articletype{ARTICLE TEMPLATE}% Specify the article type or omit as appropriate#
	
	\title{\textit{SurviveCovid-19} - An Educational Game to Facilitate Habituation of Social Distancing and Other Health Measures for \textit{Covid-19} Pandemic}
	
	\author{
		\name{Akhila Sri Manasa Venigalla\textsuperscript{a}, 
			Dheeraj Vagavolu\textsuperscript{b} and Sridhar Chimalakonda\textsuperscript{c}}
		\affil{Research in Intelligent Software \& Human Analytics (RISHA) Lab\\ Indian Institute of Technology, Tirupati, India\\ \textsuperscript{a}cs19d504@iittp.ac.in
			\textsuperscript{b}cs17b028@iittp.ac.in
			\textsuperscript{c}ch@iittp.ac.in}
	}
	
	\maketitle

	\begin{abstract}
		\textit{Covid-19} has been causing severe loss to the human race. Considering the mode of spread and severity, it is essential to make it a habit to follow various safety precautions such as using sanitizers and masks and maintaining social distancing to prevent the spread of \textit{Covid-19}. Individuals are widely educated about the safety measures against the disease through various modes such as announcements through online or physical awareness campaigns, advertisements in the media and so on. The younger generations today spend considerably more time on mobile phones and games. However, there are very few applications or games aimed to help in practicing safety measures against a pandemic, which is much lesser in the case of \textit{Covid-19}. 
		% Also, considering the majority of time spent at home in the present situation across the world, games also act as a good pass time indoors. 
		
		Hence, we propose a 2D survival-based game, \textit{SurviveCovid-19}, aimed to educate people about safety precautions to be taken for \textit{Covid-19} outside their homes by incorporating social distancing and usage of masks and sanitizers in the game. \textit{SurviveCovid-19} has been designed as an Android-based mobile game, along with a desktop (browser) version, and has been evaluated through a remote quantitative user survey, with 30 volunteers using the questionnaire based on the MEEGA+ model. The survey results are promising, with all the survey questions having a mean value greater than 3.5. The game's quality factor was 69.3, indicating that the game could be classified as excellent quality, according to the MEEGA+ model. 
		% Till here
		
	\end{abstract}
	

	\section{Introduction}
	\label{intro}
	
	Several pandemics have proved to be a risk to people across the globe during many instances. 
	The important lessons learned from pandemics witnessed by the world, such as \textit{Cholera} \citep{mutreja2011evidence, colwell1996global}, \textit{H1N1} \citep{xu2010structural}, Smallpox \citep{li2007origin, cunha2004influenza}, \textit{Ebola} \citep{richardson2016biosocial} and so on \citep{hays2005epidemics} is the role of awareness campaigns and practice of safety measures. It has been observed that public awareness campaigns reduce the severity of the outcomes of disease, including hospitalization and death \citep{mytton2012influenza}. Several governments across the world and researchers have highlighted the need to improve public health education campaigns aimed towards improving health literacy among the people \citep{eastwood2010responses}. Health professionals across various nations have been advised to practice several control activities, including basic sanitation facilities, identify and isolate suspected cases and educate the general public through awareness campaigns \citep{sepulveda2006cholera}. Along with the medical care required, all individuals worldwide must be aware of and follow safety precautions to control the pandemic \citep{lacitignola2021managing}.
	A severe outbreak of pneumonia caused by SARS-Cov-2, \textit{Covid-19} \citep{ciotti2019covid} has spread rapidly, affecting more than 2.76 million people, spread across several countries of the world, as reported by WHO as of 28 March, 2021\footnote{\url{https://www.who.int/publications/m/item/weekly-epidemiological-update-on-covid-19---31-march-2021}}. It spreads on a large scale, in a short span of time, ranging from 110.7 million cases\footnote{\url{https://www.who.int/publications/m/item/weekly-epidemiological-update---23-february-2021}} as of 23 February 2021 to 126.3 million cases as of 28 March 2021, affecting more than 15.6 million people in a span of one month (from 23 February 2021 to 28 March 2021). Symptoms of \textit{Covid-19} include fever, cough, dyspnoea, lymphocytopenia, fatigue \citep{zhou2020clinical}, \citep{yuki2020covid}, that weakens the immune system and increases complications in a few cases, leading to heart failures, kidney failures, respiratory failures, nervous system and mental health \citep{zhou2020clinical}\citep{zheng2020covid} \citep{wang2020potential}\citep{bouaziz2020vascular} \citep{hao2020psychiatric}.  The number of deaths caused due to \textit{Covid-19} have increased from 2.45M to 2.76M during 23 February 2021 to 28 March 2021 \footnote{\url{https://www.who.int/emergencies/diseases/novel-coronavirus-2019/situation-reports}}. Several measures to control the \textit{Covid-19} pandemic are being taken by WHO and various nations across the world \citep{sohrabi2020world} \citep{nussbaumer2020quarantine, guner2020covid, rozhnova2021model}. These measures include expediting the diagnosis and contact tracing process, increasing health care facilities and scaling up health care equipment, encouraging the process of discovering drugs to treat \textit{Covid-19} and bringing awareness among the public\footnote{\url{https://www.who.int/emergencies/diseases/novel-coronavirus-2019/advice-for-public}}. Vaccines are being developed to prevent the contraction of Covid-19, and ways to administer and educate about these vaccines to a wider population are being explored \citep{forni2021covid, burgess2021covid}. 
	Public Awareness programs about \textit{Covid-19} are being carried out in various ways across the globe. The public is being educated and repeatedly reminded of safety measures to be followed through telephonic and television advertisements, physical and online campaigns, flyer displays at public places, and many online websites, schools and so on \citep{rozhnova2021model}. Several websites dedicated to \textit{Covid-19} are being developed, to provide information about precautions to be taken and the status of \textit{Covid-19} \citep{venigalla2020mood}. In many countries across the world, lockdown instructions are eased, and their citizens are allowed to come outdoors. It is essential to follow safety precautions when people are in public places. Today, the younger generation spends a lot of time with their mobile phones and desktops on various apps and games \citep{chan2017adolescents}. Also, it has been observed that games can affect players' behaviour and attitude and contribute to better learning \citep{orji2013lunchtime}\citep{chen2020games}\citep{wang2020integrating}. 
	
\begin{figure}
		\centering
		\includegraphics[width = \linewidth, height = 5cm]{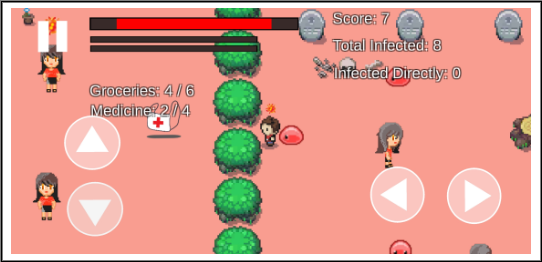}
		\caption{Snapshot of \textit{SurviveCovid-19}}
		\label{fig:snap}
	\end{figure}

	Despite the large-scale public health awareness programs being carried out, we are not aware of any mobile games aimed to educate people about the safety measures to be taken against \textit{Covid-19} and help people to make a habit of following them. Thus, we propose \textit{SurviveCovid-19}\footnote{The game is also available at \url{https://survivecovid-19.itch.io/game2020}}, a survival theme-based 2D mobile and desktop (browser) game, to educate users about various safety measures to be followed against \textit{Covid-19} when in outdoors and consequently help them in practicing these safety measures. A snapshot of the game is presented as Figure \ref{fig:snap}.
	% Till here
	
	%Also, games instigate interest among players and improve retention capacity.
	\section{Related Work}
	\label{related work}
	Bringing out awareness among public with respect to various diseases plays a primary role in keeping people healthy. With the increase in use of mobile and web applications, several health care mobile apps and websites are being developed.
	
	\subsection{Mobile Apps for Health Literacy}
	Mobile apps to support prevention of various diseases such as obesity, cardiovascular diseases, chronic diseases such as skin cancer and so on have been developed \citep{brinker2018skin}\citep{sindi2015caide}\citep{kong2012dietcam}\citep{matsumura2013iphysiometer}. Brinker et al., have presented a facial-aging mobile app, \textit{SunFace}, aimed to prevent skin cancer \citep{brinker2018skin}. \textit{SunFace} encourages users to click a selfie and view changes induced due to 5 and 25 years of skin aging with and without sun protection. It also provides information about most commonly caused skin cancers due to UV rays and displays odds ratio of skin cancer based on various behaviors \citep{brinker2018skin}. A study  conducted to evaluate \textit{SunFace} app with around 350 Brazilian secondary school students has shown increased considerations for UV protective behaviours \citep{brinker2018skin}\citep{brinker2017photoaging}. \textit{CAIDE} Dementia Risk Score App has been developed by Sindi et al. to predict risk of dementia based on the profile of individuals \citep{sindi2015caide}. It also provides guidance to individuals to reduce the risk of dementia. The ``medical history, health status and current health behaviour" of individuals is taken as input, through a questionnaire and risk score is calculated based on this information. It also provides a platform for medical practitioners to monitor risk of dementia of an individual and provide necessary guidance \citep{sindi2015caide}. 
	
	Kong et al. have developed \textit{DietCam}, a mobile phone based application that aims to reduce obesity \citep{kong2012dietcam}. It assesses the food intake of users based on three images, clicked from three different perspectives, or a short video of the food being consumed. The type of food in the image or video is classified by comparing result of feature matching with different types of food stored in food and nutrition database. The volume of the food is identified through geometry calibration and 3D reconstruction. An estimation of number of calories being consumed based on the type of food and volume generated, is presented to the users, which helps them in assessing the calorie intake, and consequently manage their diet \citep{kong2012dietcam}. Studies also reveal that more than 70\% of users, using weight loss reduction mobile applications show successful weight reduction \citep{chin2016successful}. These studies thus imply the positive outcomes of using mobile applications in health care.
	\textit{iPhysioMeter} has been proposed as a smartphone program (on iPhone 4 , 4S and 5) to measure heart rate (HR) and normalized pulse volume(NPV) \citep{matsumura2013iphysiometer}. The HR and NPV values are calculated by the program when users tap on the iPhone home screen and cover the flash LED and CMOS camera with their left index finger. The results of measurements are sent to the users' email \citep{matsumura2013iphysiometer}.   
	
	\subsection{Games for Healthcare}
	Several mobile games have also been developed to promote self health monitoring for various diseases \citep{birn2014mobilequiz}\citep{park2015snackbreaker}\citep{baghaei2016diabetic}\citep{bin2019mobile}. \textit{MobileQuiz} has been developed as a mobile game, to promote both physical and cognitive skill among older adults and is to be played outdoors \citep{birn2014mobilequiz}. It assigns tasks to users in view of the physical distance and direction to be covered by the user, which when accomplished, users are rewarded with points and are provided with a questionnaire related to the location or surrounding regions \citep{birn2014mobilequiz}. Park et al. have proposed \textit{SnackBeaker}, a mobile game, aimed to promote healthy food choices among the players \citep{park2015snackbreaker}. It presents a quiz that can be answered based on elimination of one among two choices, in terms of nutritional values provided by the two alternatives, thus educating users about the nutritional value of different snacks and consequently helping them choose snacks with high nutritional values \citep{park2015snackbreaker}. A mario style based mobile game, \textit{Diabetic mario}, was introduced to help in understanding ways to manage diabetes \citep{baghaei2016diabetic}. The player is assigned with the task of managing blood sugar level of Diabetic Mario by selectively choosing the food items provided in the game console and eventually save the princess \citep{baghaei2016diabetic}. Considering the importance of Automated External Defibrillators(AED) in cases of out-of-hospital cardiac arrests, a crowdsource based mobile game has been designed to make users vary of AED locations and also to motivate them to report condition of AEDs \citep{bin2019mobile}. 
	
	\subsection{Web-based Applications for Awareness during Pandemic}
	Considering the importance of public awareness towards a pandemic and the importance of research in view of planning and response to curb the pandemic, few online websites and telephonic lines have also been developed. ISIS has been developed as a web framework to help in planning and response against pandemics by analyzing multiple simulations of possible interventions, including ``administering a vaccine to a part of population, using antivirals as treatment, closing schools and work places and instituting social distancing" \citep{beckman2014isis}. WHO also lists out various safety measures on the WHO website to enable public awareness during pandemics. Few web based applications and websites are also being developed to help in tracking \textit{Covid-19} reported cases and in bringing public awareness towards \textit{Covid-19}. Dong et al have developed web-based dashboard that displays visualizations of \textit{Covid-19} reported cases from time to time \citep{dong2020interactive}. It ``reports cases of the \textit{Covid-19} pandemic at level of cities in USA, Australia and Canada, province level in China and country level in other cases" \citep{dong2020interactive}. Few governments across the world have also promoted use of various mobile based applications\footnote{\url{http://tiny.cc/dxqfnz}}\footnote{\url{http://tiny.cc/bzqfnz}}\footnote{\url{http://tiny.cc/70qfnz}}\footnote{\url{http://tiny.cc/t1qfnz}} that are aimed to bring awareness among people about risks, best practices and relevant advisories regarding \textit{Covid-19}. 
	
	Considering the positive impact of mobile and desktop games and apps \citep{orji2013lunchtime} in the area of health care and the proven usefulness in bringing awareness about various diseases and inculcating healthy habits through mobile games and apps, designing  games to help in public awareness in the context of \textit{Covid-19} pandemic could be useful. However, inspite of the availability of mobile and web applications towards containment of \textit{Covid-19}, we are not aware of any mobile or desktop games that promote inculcating the habit of following safety measures with respect to \textit{Covid-19}. Hence, we present \textit{SurviveCovid-19}, a survival based 2D game, aimed to educate users and help in practicing various safety measures to be followed against \textit{Covid-19} pandemic. 
	
	\section{Design}
	\label{design}

	% \textit{Covid-19} is severely affecting various nations across the world, at a rapid rate in the present day. It has been observed to spread through physical contact and respiratory droplets. 
	% Hence, several countries have issued lockdown instructions so as to prevent public gatherings and consequently, flatten the curve. 
	% Maintaining physical distance, using masks to cover nose and mouth and frequently using sanitizers to keep palms clean and visiting a doctor when symptoms of cough and fever are observed, help in reducing the risk of contracting the virus. It is important to inculcate the habit of following the measures mentioned above among the public.
	Multiple studies in the literature propose various ways to improve effectiveness and simplify development of serious games for educational purposes, which included sets of guidelines based on the target audience for the games \citep{valenza2019serious, westera2019and, manuel2019simplifying}. The guidelines presented in \citep{westera2019and} were observed to be more closely related to the goals of \textit{SurviveCovid-19}, such as motivation and reduced cognitive load. 
	
	\textit{SurviveCovid-19} has thus been designed keeping in mind the desirable factors for effectiveness of serious games for better learning outcomes, such as better motivation and reduced cognitive load, as suggested in \citep{westera2019and}. This is an educational game that helps people in understanding the importance of masks and sanitizers and following the safety measures to keep themselves and people in their surroundings safe from \textit{Covid-19}. This game is inspired by a simple pixel-based top-down style design where people navigate the city with safety and health measures.

	The game begins by showing a landing screen. As the game starts, the player sees a short video depicting a simple story-line of a person who needs to get essential items from the city during lock down for his family while following proper health measures and avoid spreading the virus. The scenes displayed in the game are designed to be appealing to the users, and also resemble real-life scenarios, thus meeting the styling criteria and reducing cognitive load, as mentioned in \citep{westera2019and}. The video is then followed by a screen showing the specific instructions to complete the game. The goal of the player is to collect the allotted number of groceries and medicines while navigating the city. While doing so, the player also has to make use of face masks and sanitizers to avoid getting infected or infecting other people. To make the game challenging and fun for the players, both of these utilities have been given a timer. The player has to make use of them strategically to win. The game finishes when the player collects all the required items and reaches home, as shown on the map. 
	
	The game also keeps in check the number of people getting affected because of the user's action in terms of people being infected directly and indirectly. This prompts the player to tread the path with caution and by taking proper health measures throughout the game. The game rewards the player with safety shield when masks or sanitizers are used with-in the game, thus motivating the players to make use of these elements.

	\begin{figure}
		\centering
		\includegraphics[width = \linewidth]{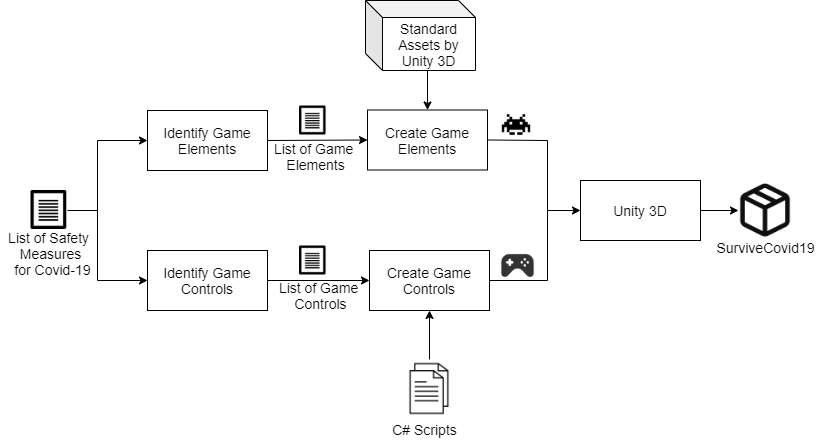}
		\caption{Development of \textit{SurviveCovid-19}}
		\label{fig:dev}
	\end{figure}

	\section{Development}
	\label{dev}

	We have developed \textit{SurviveCovid}, as an motivational and educational game to learn the importance of, and to follow proper health measures against \textit{Covid-19} using the Unity 3D game engine\footnote{\url{https://unity.com/}}. We followed a step by step approach to develop the game as shown in Figure \ref{fig:dev}. We then used the Android SDK module provided by Unity to export the game for android based mobile devices.
	
	We have used 2d tiles from the standard assets pack provided by Unity to define most of the game objects. The rest have been created using Photoshop software. The tiles are first imported into Unity and then using the tile-map feature of Unity, we design the game. In this way, the map of the city in which the player has to navigate is created. A mask and a sanitizer object are created and have been placed at strategic positions on the map. These utilities are meant to help the player to complete the level. 
	
	All the events in the game are controlled using scripts written in \textit{C\#} programming language. A mechanism to simulate the spread of disease through touch has been implemented using \textit{C\#} scripts. This spread can be prevented by taking appropriate measures like using the mask and the sanitizer. 
	
	Many obstacles, such as infected people and monsters which spread the virus on touch, are placed on the map. The people are positioned such that they mimic real crowds and queues of people.
	
	A single script keeps track of the groceries and medicine that have been collected by the player. A Quota is assigned to the player automatically to collect groceries and medicine. Only on completing the assigned quota, the player may cross the finish line.
	
	All the media, such as videos and photos in the game, are made using Windows Video Maker and Photoshop softwares.

	\section{User Scenario}
	\label{userScenario}

	\begin{figure}
		\centering
		\includegraphics[width = \linewidth, height = 5cm]{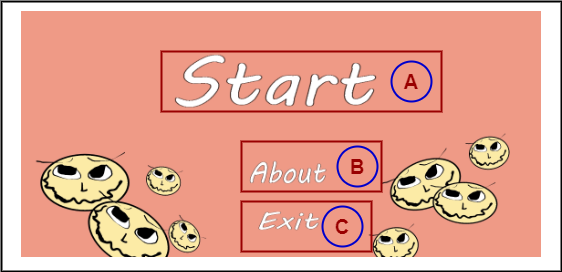}
		\caption{Start screen of \textit{SurviveCovid-19} depicting [A] Start option, [B] About option and [C] Exit option}
		\label{fig:us1}
	\end{figure}
	
	\begin{figure}
		\centering
		\includegraphics[height = 5cm, width = \linewidth]{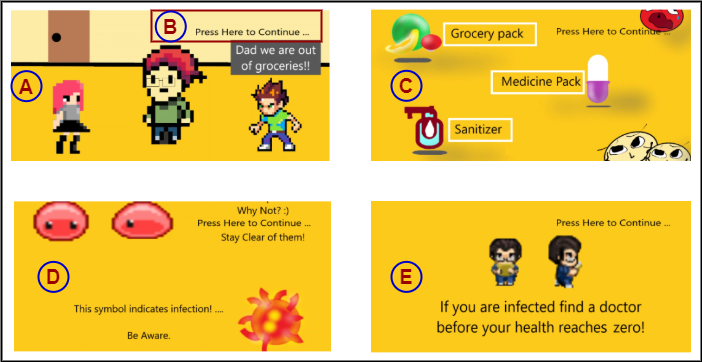}
		\caption{Snapshots of Short Video describing different game elements including [A], conversation in video, [B] option to skip video, [C] Grocery, medicine pack and sanitizer, [D] virus in the game and [E] doctor }
		\label{fig:us2}
	\end{figure}
	
	\begin{figure}
		\centering
		\includegraphics[width = \linewidth]{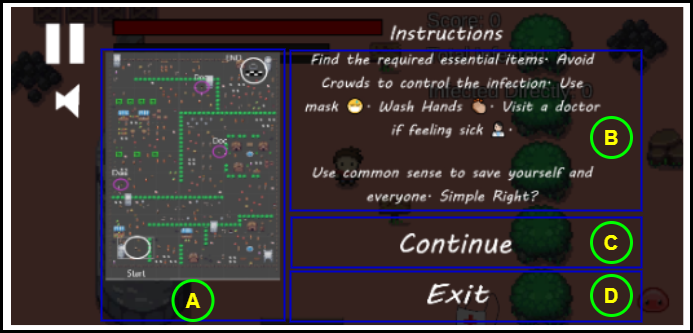}
		\caption{Instruction page of \textit{SurviveCovid-19} depicting [A] Map of the city in the game, [B] Instructions in the game, [C] Option to continue and [D] Option to Exit the game}
		\label{fig:us3}
	\end{figure}
	
	\begin{figure}
		\centering
		\includegraphics[width = \linewidth]{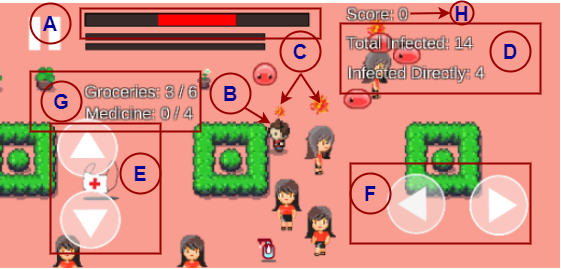}
		\caption{Snapshot of \textit{SurviveCovid-19} depicting [A] Decreasing lifeline of player effected with virus, [B] Player effected with Virus, [C] Player infecting other individual, [D] Count of infected individuals, [E] Controls to navigate up and down [F] Controls to navigate left and right, [G] Information about Groceries and Medicine collected vs to be collected and [H] Score of the player }
		\label{fig:us4}
	\end{figure}
	
	\begin{figure}
		\centering
		\includegraphics[width = \linewidth]{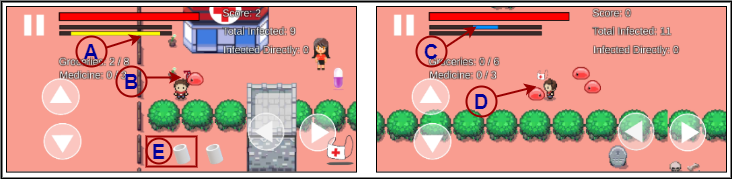}
		\caption{Snapshots of \textit{SurviveCovid-19} depicting [A] Lifeline of sanitizer, [B] Player not being effected by virus in presence of sanitizer, [C] Lifeline of mask, [D] Player not being effected by virus in presence of mask and [E] Game elements that contribute to increase in Player's score}
		\label{fig:us5}
	\end{figure}
	
	\begin{figure}
		\centering
		\includegraphics[width = \linewidth]{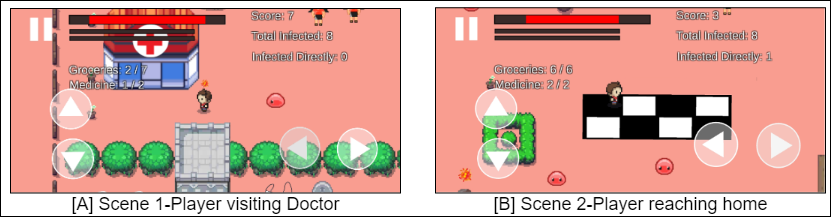}
		\caption{Snapshots of \textit{SurviveCovid-19} depicting [A] Infected player visiting a doctor and [B] Player reaching home on accomplishing tasks of Groceries and Medicine}
		\label{fig:us6}
	\end{figure}
	Consider \textit{Moksha}, an individual staying indoors due to lockdown instructions in her locality, interested to play games and to gain awareness against \textit{Covid-19}. She considers playing \textit{SurviveCovid-19}, to pass her time and also to be habituated to follow safety measures when in outdoors. She downloads \textit{SurviveCovid-19} on her Android mobile and starts the game. She is displayed with a start screen, with options - Start, About and Exit as shown in Figure \ref{fig:us1}. She selects to start the game by clicking on [A] of Figure \ref{fig:us1}. She is then displayed with a short video that describes the lockdown situation of a person, who is required to go outdoors to get daily essentials and medicine. \textit{Moksha} can also choose to skip the video and directly navigate to the game by clicking on \textit{Press here to Continue} ([B] of Figure \ref{fig:us2}),  A snapshot of the video is shown in [A] of Figure \ref{fig:us2}. Information about various game elements is also displayed as a part of the video, as shown in [C] of Figure \ref{fig:us2}. \textit{Moksha} is then presented with Instructions screen containing instructions to be followed as shown in [A] of Figure \ref{fig:us3}, a map depicting layout of the city, shown in [B] Figure \ref{fig:us3}, options to continue and exit, as shown in [C] and [D] of Figure \ref{fig:us3} respectively. She clicks on \textit{Continue} ([C] of Figure \ref{fig:us3}) and starts the game. 
	
	\textit{Moksha} is then displayed with the number of Groceries and Medicines to be collected as shown in [G] of Figure \ref{fig:us4} The number of groceries and medicines collected versus the goal to be collected are displayed in [G] of Figure \ref{fig:us4}. [A] of Figure \ref{fig:us4} depicts lifeline of the player, while the score is displayed to the top right, as shown in [H] of Figure \ref{fig:us4}. The player can move around in the city using the controls depicted in [E] and [F] of Figure \ref{fig:us4}.
	In cases where the player is infected with virus and gets in contact with other individuals in the game, the player infects those individuals as shown in [C] of Figure \ref{fig:us4}. When these infected individuals move around the city, they also infect other healthy individuals. The total number of individuals infected by the player directly and indirectly is represented in [D] of Figure \ref{fig:us4}.
	She starts moving around the city to collect required essentials. Considering the safety precautions, not to be affected by the virus, she collects the mask and sanitizer wherever available. The lifeline of existence of sanitizer is displayed below the player's lifeline, as depicted in [A] of Figure \ref{fig:us5}. [B] of Figure \ref{fig:us5} indicates virus not being able to harm player in the presence of sanitizer. The lifeline of mask is depicted in [C] of Figure \ref{fig:us5} and [D] indicates the mask protecting player from the virus. Score of the player increases on collecting elements displayed in [E] of Figure \ref{fig:us6}. Player contracts the virus in absence of mask or sanitizer, thus resulting in declining lifeline of the player. \textit{Moksha} then visits a doctor as shown in [A] of Figure \ref{fig:us6}, to get cured and revive player's lifeline. Once all the groceries and medicine are collected, player reaches home, as shown in [B] of Figure \ref{fig:us6}.
	
	\section{Evaluation}
	\label{Eval}
	\textit{SurviveCovid-19} has been developed to educate about, and motivate public to follow safety measures to be taken to control \textit{Covid-19} pandemic, as an Android and Web based mobile game. Though we could not find games developed to support public health awareness towards pandemics, several health care-related and learning-oriented games have been evaluated based on similar factors such as learning outcomes, usability, motivation, user experience, usefulness and so on, based on different evaluation models such as TAM, MEEGA+ and so on. % Also, several educational games have been evaluated based on usability and player experience. 
	
	Among the existing evaluation models, we see that MEEGA+ has been commonly used to evaluate educational games \citep{santos2019risking, tsopra2020antibiogame}, as it considers desired factors such as fun, challenge, learnability, operability and so on, thus including both player experience and usability. As these factors are also relevant for the evaluation of \textit{SurviveCovid-19}, we adapt MEEGA+ for evaluating the game. However, as \textit{SurviveCovid-19} aims to influence players to follow safety precautions against Covid-19, we also evaluate its influence on players towards making a habit to follow the safety measures. Also, we design \textit{SurviveCovid-19} based on the suggestions in \citep{westera2019and}, to make it more effective for learning. Thus, we evaluate \textit{SurviveCovid-19} based on the factors - intrinsic motivation and reduced cognitive load. Hence, we design the questionnaire based on an amalgamation of the factors discussed above. The factors considered for evaluation are as follows:
	\begin{itemize}
		\item Influence/Impact (I)
		\item In-game Safety Precautions (SP) 
		\item Motivation 
		\begin{itemize}
			\item Competence (C)
			\item Autonomy (A)
			\item Style Elements (SE)
			\item Rewards (RW)
		\end{itemize}
		\item Reduced Cognitive Load (RCL)
		\item Player Experience
		\begin{itemize}
			\item Focused Attention (FA)
			\item Satisfaction (S)
			\item Relevance (R)
			\item Fun (F)
			\item Challenge (Ch)
		\end{itemize}
		\item Usability
		\begin{itemize}
			\item Learnability (L)
			\item Operability (O)
			\item Aesthetics (As)
			\item Accessibility (Acc)
		\end{itemize}
	\end{itemize}
	
	The current version of \textit{SurviveCovid-19} is a single-player game, targeted at a wider audience, outside the classroom. Hence, some factors of MEEGA+ such as Social Interaction, Perceived Learning and User Error Protection have been omitted for evaluation of the game. We design a questionnaire, as presented in Table \ref{tab:table1} based on the factors mentioned above.
	% Hence, we have evaluated \textit{SurviveCovid-19} to understand its usefulness, influence, usability and player experience. 
	
	Emails were sent out to 70 individuals with varied professional background, which included doctors, architects, engineers, chartered accountants, and students pursuing medicine, engineering and architecture. The email consisted of information about the game, link to browser and android versions of the game, and link to the user survey.  
	These individuals were requested to either download the \textit{SurviveCovid-19} application and install the same on their Android mobile phones or navigate to the browser version of the game. A 5-point Likert Scale based questionnaire as shown in Table \ref{tab:table1}, has been sent to all the participants as a user survey in the email. Apart from the questions based on the factors considered, as presented in  Table \ref{tab:table1}, other questions that could support the analysis such as demographic questions were also included in the user survey (as presented in Table \ref{tab:demographics}). The participants were asked to play the game and provide their feedback by answering the questionnaire presented in the user survey. Of these 70 individuals, 30 individuals responded to the user survey after playing the game. 
	
	% We considered 30 volunteers in the age group of 17-27 years and conducted a remote quantitative user survey based on the questionnaire in Table \ref{tab:table1}. 
	
	\begin{table}[]
		\caption{Quantitative User Survey.}
		\label{tab:table1}
		\begin{tabular}{|l|l|l|l|}
			\hline
			\textbf{Factors} & \textbf{Questions}                                                                                                     & \textbf{Mean} & \textbf{SD} \\ \hline
			I                & The theme of the game influenced my actions in real-time                                                               & 3.67          & 1.09        \\ \hline
			I                & After playing the game for more than 3 times, I intended      & 3.5           & 1.25        \\ 
			& to use sanitizers more frequently in my day-to-day life.      &           &        \\ \hline
			
			I                & After playing the game for more than 3 times, I intended & 3.67          & 1.12     
			\\
			&  to follow  more often social distancing in my day-to-day life & & \\
			\hline
			I                & After playing the game for more than 3 times, I intended            & 3.64          & 1.16        \\ 
			& to more frequently use masks in my day-to-day life & & \\\hline
			RW/SP            & There were in-game rewards that motivated me                                   & 3.87          & 1.1         \\ 
			& to follow wearing masks in the game     & & \\\hline
			RW/SP            & There were in-game rewards that motivated me                               & 3.67          & 1.09        \\
			& to follow social distancing in the game     & & \\\hline
			RW/SP            & There were in-game rewards that motivated me                                 & 3.76          & 1.22        \\
			& to follow using sanitizers in the game    & &\\\hline
			C                & When I first looked at the game, I had the                               & 3.9           & 1.02        \\
			& impression that it would be easy for me.      & &\\
			\hline
			A                & It is due to my personal effort that I                                         & 3.73          & 0.94        \\
			&managed to advance in the game.         & &\\
			\hline
			A                & I could reach the target in the game on my own                                                                         & 3.76          & 1           \\ \hline
			SE/Acc           & The fonts (size and style) used in the game are easy to read.                                                          & 4             & 0.83        \\ \hline
			SE/As             & The game design is attractive (interface, graphics, etc.).                                              & 4.06          & 0.98        \\ \hline
			RCL              & Game presents visuals of real-time scenarios, which helped                     & 3.76          & 1.04        \\ 
			& me to relate well to my surroundings     & &
			
			\\
			\hline
			FA               & There was something interesting at the beginning                            & 3.8           & 1.03        \\
			&  of the game that captured my attention.   & &\\
			\hline
			S                & I would recommend this game to my colleagues.                                                                          & 4.1           & 0.84        \\ \hline
			S                & I feel satisfied with the things that I learned from the game.                                                         & 3.8           & 1.03        \\ \hline
			S                & Completing the game tasks gave me a satisfying                                         & 3.9           & 0.88        \\ 
			&feeling of accomplishment.      & &\\
			\hline
			R                & The game contents are relevant to my interests.                                                                        & 3.67          & 1.09        \\ \hline
			F                & Something happened during the game (game elements,                         & 3.93          & 0.98        \\
			&competition, etc.) which made me smile.     & &\\
			\hline
			F                & I had fun with the game.                                                                                               & 4.03          & 0.8         \\ \hline
			Ch               & The game does not become monotonous as it progresses                        & 3.67          & 1.02        \\
			&(repetitive or boring tasks).              & &\\
			\hline
			Ch               & The game provides new challenges (offers new obstacles,             & 3.64          & 1.15        \\
			&situations or variations) at an appropriate pace.  & &\\
			\hline
			Ch               & This game is appropriately challenging for me.                                                                         & 3.84          & 0.94        \\ \hline
			L                & I think that most people would learn to play                                                    & 3.87          & 0.97        \\ 
			&this game very quickly.& &\\
			\hline
			L                & Learning to play this game was easy for me.                                                                            & 3.87          & 0.97        \\ \hline
			O                & The game rules are clear and easy to understand.                                                                       & 4             & 0.94        \\ \hline
			O                & I think that the game is easy to play.                                                                                 & 3.93          & 0.82        \\ \hline
		\end{tabular}
	\end{table}
	
	\begin{table}[]
		\centering
		\label{tab:demographics}
		\caption{Demographics}
		\begin{tabular}{|l|r|}
			\hline
			\multicolumn{2}{|l|}{\textbf{Gender}}                                                  \\ \hline
			Female                                         & 13                                    \\ \hline
			Male                                           & 17                                    \\ \hline
			
			\multicolumn{2}{|l|}{\textbf{Age (in years)}}                                          \\ \hline
			17-20                                          & 6                                     \\ \hline
			21-24                                          & 17                                    \\ \hline
			24-27                                          & 7                                     \\ \hline
			\multicolumn{2}{|l|}{\textbf{Profession}}                                              \\ \hline
			Architect                                      & 3                                     \\ \hline
			Architecture Student                           & 2                                     \\ \hline
			Doctor                                         & 2                                     \\ \hline
			Public Health/Medical Student                  & 2                                     \\ \hline
			Software Engineer                              & 4                                     \\ \hline
			Engineering Student                            & 16                                    \\ \hline
			CA                                             & 1                                     \\ \hline
			\multicolumn{2}{|l|}{\textbf{Frequency of Playing Games}}                            \\ \hline
			Very Often                                     & 8                                     \\ \hline
			Often                                          & 6                                     \\ \hline
			Neutral                                        & 1                                     \\ \hline
			Sometimes                                      & 10                                    \\ \hline
			Never                                          & 5                                     \\ \hline
			
			\multicolumn{2}{|l|}{\textbf{Version Played}} \\ \hline
			Mobile                                         & 20                                    \\ \hline
			Desktop/Browser                                & 3                                     \\ \hline
			Both                                           & 7                                     \\ \hline
		\end{tabular}
		\vspace{-1cm}
	\end{table}

	\section{Results}
	\label{results}
	
	\begin{figure}
		\centering
		\includegraphics[width = \linewidth]{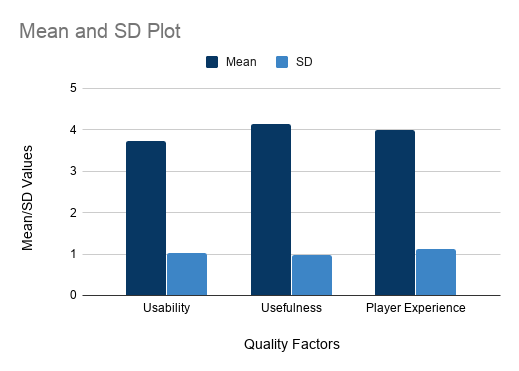}
		\caption{Result of Questionnaire in terms of Mean and Standard Deviation of Quality Factors}
		\label{fig:results}
	\end{figure}
	
	We observed that there were 13 female and 17 male participants among the 30 volunteers who responded to the user survey (these details are presented in Table \ref{tab:demographics}). These participants had varied professional backgrounds, which also included 2 doctors, 2 public health and medical students, 2 architects, 2 architecture students and 16 engineering students, apart from software engineers and chartered accountants. Majority of the volunteers belonged to the age group of 21-24 years and 25 participants have agreed to have played games at least sometimes. All the participants have played the \textit{SurviveCovid-19} game on an average of 5 times, before answering the user survey questionnaire.
	
	Table \ref{tab:table1} indicates the results of user survey in terms of mean and standard deviation of each question in the questionnaire. It can be observed that the mean for all the questions is greater than or equal to 3.5, indicating the positive experience of users. 
	Figure \ref{fig:results}, depicts the mean and standard deviation values for each of the three quality factors. The mean values for each of the quality factors were considered to calculate the quality score based on MEEGA+ model. Applying the Cronbach Alpha value for means of player experience (FA, F, Ch, S, R) and usability (L, O, As, Acc) factors, and then normalizing the resultant value by multiplying it by 10 resulted in quality score of 69.3. The MEEGA+ evaluation theory states that games with quality score greater than 65 are of excellent quality, while that between 42.5 to 65 are considered to be of better quality. The quality score of \textit{SurviveCovid-19} is 69.3, indicating that the game could be included in excellent quality category. This result indicates that such games are challenging for learners, without monotonous activities and clear rules to play the game. The mean of motivation factors (C, A, SE, RW), and other factors (I, SP, RCL) is 3.76, indicating that majority of the participants find \textit{SurviveCovid-19} to be motivating and helps the players in inculcating safety precautions against Covid-19 as a habit.
	
	%  Mean value of 3.7 for Usability indicates that most of the volunteers found \textit{SurviveCovid-19} easy and attractive. The mean value for Usefulness is greater than 4, indicating that most of the volunteers intended to follow the safety measures against \textit{Covid-19} while playing the game. Player Experience factor having mean value greater than 3.9 reveals that majority of the volunteers found \textit{SurviveCovid-19} to be interesting to play and were willing to recommend \textit{SurviveCovid-19} to their peers. 
	Participants also suggested that :
	
	\begin{itemize}
		\item \textit{Control sensitivity for navigation buttons could be improved.}
		\item \textit{Way to home and to hospitals could be indicated in the game.}
		\item \textit{The game button reaction is quite fast , rest all is fine}
	\end{itemize}

	\section{Discussion and Limitations}
	\label{limit}
	
	\textit{SurviveCovid-19} is based on survival theme, where the storyline deals with a person who needs to go outdoors to purchase daily essentials and medicine during \textit{Covid-19} situation. The game has been designed in a way which enforces the players to follow safety measures to stay safe from \textit{Covid-19} when outdoors. Thus, the developed prototype version of \textit{SurviveCovid-19} aims to educate players about precautions to be taken and also inculcate the habit of following these precautions against \textit{Covid-19}. 
	% and also serves as a pass time during the lockdown situations that are prevailed due to \textit{Covid-19}. 
	Though this game has been developed to address the younger generations of society, it can be played by a wide range audience, irrespective of any social or educational backgrounds. The only requirement of the game is that the players are expected to be equipped with using a mobile phone or a desktop/laptop.
	
	The current version of \textit{SurviveCovid-19} has only one level. However, further levels can easily be added to the game and difficulty level could also be increased with every level being added. The game could also include roles for the player, such as scientist, doctor and health-care workers, and also include the aspect of vaccines. Also, the existing \textit{SurviveCovid-19} application can be played only on Android based mobile devices and on browser (from desktop). The future versions of the game could be made compatible to mobile devices operating on other operating systems such as Apple and Windows.

	%single player and not online
	
	\section{Conclusion and Future Work}
	\label{conclusion}
	In this paper, we presented the prototype version of \textit{SurviveCovid-19}, a survival based 2D mobile and desktop (browser) game, aimed to bring awareness about safety measures to be followed against \textit{Covid-19}. Public Awareness Campaigns are considered as one of the most important measures to be taken across the world in cases of a pandemic or a disease outbreak \citep{mytton2012influenza}, \citep{eastwood2010responses}. The current outbreak of \textit{Covid-19} pandemic requires all individuals across the world to follow precautions to prevent the spread of \textit{Covid-19}. Large scale public health awareness campaigns are being carried out by almost all the governments and various organizations across the world, including WHO to control \textit{Covid-19}. Considering the large amount of time spent by younger generations on mobile and desktop apps and games in the present day, and the effect of games on players' attitude \citep{chan2017adolescents}, \citep{orji2013lunchtime}, it could be useful to introduce games that contribute towards public health awareness and help the public in inculcating safety precautions as a habit. \textit{SurviveCovid-19} aims to educate people about the safety measures to be followed during \textit{Covid-19} and motivates them to follow the precautions in real time. We have evaluated \textit{SurviceCovid-19} through a questionnaire adapted from factors in MEEGA+ model, and other factors such as motivation and impact of the game. The evaluation was done through a remote quantitative user survey with 30 volunteers, and obtained positive feedback with game quality score equal to 69.3, indicating excellent quality category of the game. Majority  of the volunteers found the game to be motivating and interesting (more than 70\% responded with agree or strongly agree for the respective questions).

	The game currently supports Android mobile devices. It can further be extended to support mobile devices running on other operating systems. We plan to improve the game by increasing number of levels in the game and difficulty of the levels.  Also, future versions of the game can include online and multiplayer features unlike the current offline and single player version of the game. We also plan to modify the game controls by improving the sensitivity of the control keys, as suggested by the participants in the survey.

	\section*{Acknowledgement(s)}
	We would like to thank Richard Lounsbery Foundation for their support and encouragement. We thank all the volunteers for their valuable time and honest feedback.

	\bibliographystyle{apacite}
	\bibliography{interactapasample}

\begin{thebibliography}{}

\bibitem [\protect \citeauthoryear {%
Baghaei%
, Nandigam%
, Casey%
, Direito%
\BCBL {}\ \BBA {} Maddison%
}{%
Baghaei%
\ \protect \BOthers {.}}{%
{\protect \APACyear {2016}}%
}]{%
baghaei2016diabetic}
\APACinsertmetastar {%
baghaei2016diabetic}%
\begin{APACrefauthors}%
Baghaei, N.%
, Nandigam, D.%
, Casey, J.%
, Direito, A.%
\BCBL {}\ \BBA {} Maddison, R.%
\end{APACrefauthors}%
\unskip\
\newblock
\APACrefYearMonthDay{2016}{}{}.
\newblock
{\BBOQ}\APACrefatitle {Diabetic Mario: Designing and evaluating mobile games
  for diabetes education} {Diabetic mario: Designing and evaluating mobile
  games for diabetes education}.{\BBCQ}
\newblock
\APACjournalVolNumPages{Games for health journal}{5}{4}{270--278}.
\PrintBackRefs{\CurrentBib}

\bibitem [\protect \citeauthoryear {%
Beckman%
\ \protect \BOthers {.}}{%
Beckman%
\ \protect \BOthers {.}}{%
{\protect \APACyear {2014}}%
}]{%
beckman2014isis}
\APACinsertmetastar {%
beckman2014isis}%
\begin{APACrefauthors}%
Beckman, R.%
, Bisset, K\BPBI R.%
, Chen, J.%
, Lewis, B.%
, Marathe, M.%
\BCBL {}\ \BBA {} Stretz, P.%
\end{APACrefauthors}%
\unskip\
\newblock
\APACrefYearMonthDay{2014}{}{}.
\newblock
{\BBOQ}\APACrefatitle {Isis: A networked-epidemiology based pervasive web app
  for infectious disease pandemic planning and response} {Isis: A
  networked-epidemiology based pervasive web app for infectious disease
  pandemic planning and response}.{\BBCQ}
\newblock
\BIn{} \APACrefbtitle {Proceedings of the 20th ACM SIGKDD international
  conference on Knowledge discovery and data mining} {Proceedings of the 20th
  acm sigkdd international conference on knowledge discovery and data mining}\
  (\BPGS\ 1847--1856).
\PrintBackRefs{\CurrentBib}

\bibitem [\protect \citeauthoryear {%
Bin~Hussein%
\ \protect \BOthers {.}}{%
Bin~Hussein%
\ \protect \BOthers {.}}{%
{\protect \APACyear {2019}}%
}]{%
bin2019mobile}
\APACinsertmetastar {%
bin2019mobile}%
\begin{APACrefauthors}%
Bin~Hussein, M\BPBI I\BPBI H.%
, Fong, J\BPBI H.%
, Lim, C\BPBI X.%
, Lee, J\BPBI S.%
, Tan, C\BPBI T.%
\BCBL {}\ \BBA {} Ng, Y\BPBI Y.%
\end{APACrefauthors}%
\unskip\
\newblock
\APACrefYearMonthDay{2019}{}{}.
\newblock
{\BBOQ}\APACrefatitle {Mobile Application for Crowdsourced Gamification of
  Automated External Defibrillator (AED) Locations} {Mobile application for
  crowdsourced gamification of automated external defibrillator (aed)
  locations}.{\BBCQ}
\newblock
\BIn{} \APACrefbtitle {Proceedings of the 4th International Workshop on
  Multimedia for Personal Health \& Health Care} {Proceedings of the 4th
  international workshop on multimedia for personal health \& health care}\
  (\BPGS\ 24--31).
\PrintBackRefs{\CurrentBib}

\bibitem [\protect \citeauthoryear {%
Birn%
, Holzmann%
\BCBL {}\ \BBA {} Stech%
}{%
Birn%
\ \protect \BOthers {.}}{%
{\protect \APACyear {2014}}%
}]{%
birn2014mobilequiz}
\APACinsertmetastar {%
birn2014mobilequiz}%
\begin{APACrefauthors}%
Birn, T.%
, Holzmann, C.%
\BCBL {}\ \BBA {} Stech, W.%
\end{APACrefauthors}%
\unskip\
\newblock
\APACrefYearMonthDay{2014}{}{}.
\newblock
{\BBOQ}\APACrefatitle {MobileQuiz: A serious game for enhancing the physical
  and cognitive abilities of older adults} {Mobilequiz: A serious game for
  enhancing the physical and cognitive abilities of older adults}.{\BBCQ}
\newblock
\BIn{} \APACrefbtitle {International Conference on Universal Access in
  Human-Computer Interaction} {International conference on universal access in
  human-computer interaction}\ (\BPGS\ 3--14).
\PrintBackRefs{\CurrentBib}

\bibitem [\protect \citeauthoryear {%
Bouaziz%
\ \protect \BOthers {.}}{%
Bouaziz%
\ \protect \BOthers {.}}{%
{\protect \APACyear {2020}}%
}]{%
bouaziz2020vascular}
\APACinsertmetastar {%
bouaziz2020vascular}%
\begin{APACrefauthors}%
Bouaziz, J.%
, Duong, T.%
, Jachiet, M.%
, Velter, C.%
, Lestang, P.%
, Cassius, C.%
\BDBL {}others%
\end{APACrefauthors}%
\unskip\
\newblock
\APACrefYearMonthDay{2020}{}{}.
\newblock
{\BBOQ}\APACrefatitle {Vascular skin symptoms in COVID-19: a french
  observational study} {Vascular skin symptoms in covid-19: a french
  observational study}.{\BBCQ}
\newblock
\APACjournalVolNumPages{Journal of the European Academy of Dermatology and
  Venereology}{34}{9}{e451--e452}.
\PrintBackRefs{\CurrentBib}

\bibitem [\protect \citeauthoryear {%
Brinker%
\ \protect \BOthers {.}}{%
Brinker%
\ \protect \BOthers {.}}{%
{\protect \APACyear {2018}}%
}]{%
brinker2018skin}
\APACinsertmetastar {%
brinker2018skin}%
\begin{APACrefauthors}%
Brinker, T\BPBI J.%
, Heckl, M.%
, Gatzka, M.%
, Heppt, M\BPBI V.%
, Rodrigues, H\BPBI R.%
, Schneider, S.%
\BDBL {}others%
\end{APACrefauthors}%
\unskip\
\newblock
\APACrefYearMonthDay{2018}{}{}.
\newblock
{\BBOQ}\APACrefatitle {A skin cancer prevention facial-aging mobile app for
  secondary schools in Brazil: appearance-focused interventional study} {A skin
  cancer prevention facial-aging mobile app for secondary schools in brazil:
  appearance-focused interventional study}.{\BBCQ}
\newblock
\APACjournalVolNumPages{JMIR mHealth and uHealth}{6}{3}{e60}.
\PrintBackRefs{\CurrentBib}

\bibitem [\protect \citeauthoryear {%
Brinker%
\ \protect \BOthers {.}}{%
Brinker%
\ \protect \BOthers {.}}{%
{\protect \APACyear {2017}}%
}]{%
brinker2017photoaging}
\APACinsertmetastar {%
brinker2017photoaging}%
\begin{APACrefauthors}%
Brinker, T\BPBI J.%
, Schadendorf, D.%
, Klode, J.%
, Cosgarea, I.%
, R{\"o}sch, A.%
, Jansen, P.%
\BDBL {}Izar, B.%
\end{APACrefauthors}%
\unskip\
\newblock
\APACrefYearMonthDay{2017}{}{}.
\newblock
{\BBOQ}\APACrefatitle {Photoaging mobile apps as a novel opportunity for
  melanoma prevention: pilot study} {Photoaging mobile apps as a novel
  opportunity for melanoma prevention: pilot study}.{\BBCQ}
\newblock
\APACjournalVolNumPages{JMIR mHealth and uHealth}{5}{7}{e101}.
\PrintBackRefs{\CurrentBib}

\bibitem [\protect \citeauthoryear {%
Burgess%
\ \protect \BOthers {.}}{%
Burgess%
\ \protect \BOthers {.}}{%
{\protect \APACyear {2021}}%
}]{%
burgess2021covid}
\APACinsertmetastar {%
burgess2021covid}%
\begin{APACrefauthors}%
Burgess, R\BPBI A.%
, Osborne, R\BPBI H.%
, Yongabi, K\BPBI A.%
, Greenhalgh, T.%
, Gurdasani, D.%
, Kang, G.%
\BDBL {}others%
\end{APACrefauthors}%
\unskip\
\newblock
\APACrefYearMonthDay{2021}{}{}.
\newblock
{\BBOQ}\APACrefatitle {The COVID-19 vaccines rush: participatory community
  engagement matters more than ever} {The covid-19 vaccines rush: participatory
  community engagement matters more than ever}.{\BBCQ}
\newblock
\APACjournalVolNumPages{The Lancet}{397}{10268}{8--10}.
\PrintBackRefs{\CurrentBib}

\bibitem [\protect \citeauthoryear {%
Chan%
, Kow%
\BCBL {}\ \BBA {} Cheng%
}{%
Chan%
\ \protect \BOthers {.}}{%
{\protect \APACyear {2017}}%
}]{%
chan2017adolescents}
\APACinsertmetastar {%
chan2017adolescents}%
\begin{APACrefauthors}%
Chan, A.%
, Kow, R.%
\BCBL {}\ \BBA {} Cheng, J\BPBI K.%
\end{APACrefauthors}%
\unskip\
\newblock
\APACrefYearMonthDay{2017}{}{}.
\newblock
{\BBOQ}\APACrefatitle {Adolescents' perceptions on smartphone applications
  (apps) for health management} {Adolescents' perceptions on smartphone
  applications (apps) for health management}.{\BBCQ}
\newblock
\APACjournalVolNumPages{Journal of Mobile Technology in
  Medicine}{6}{2}{47--55}.
\PrintBackRefs{\CurrentBib}

\bibitem [\protect \citeauthoryear {%
Chen%
, Zhang%
, Qi%
\BCBL {}\ \BBA {} Yang%
}{%
Chen%
\ \protect \BOthers {.}}{%
{\protect \APACyear {2020}}%
}]{%
chen2020games}
\APACinsertmetastar {%
chen2020games}%
\begin{APACrefauthors}%
Chen, S.%
, Zhang, S.%
, Qi, G\BPBI Y.%
\BCBL {}\ \BBA {} Yang, J.%
\end{APACrefauthors}%
\unskip\
\newblock
\APACrefYearMonthDay{2020}{}{}.
\newblock
{\BBOQ}\APACrefatitle {Games Literacy for Teacher Education} {Games literacy
  for teacher education}.{\BBCQ}
\newblock
\APACjournalVolNumPages{Educational Technology \& Society}{23}{2}{77--92}.
\PrintBackRefs{\CurrentBib}

\bibitem [\protect \citeauthoryear {%
Chin%
\ \protect \BOthers {.}}{%
Chin%
\ \protect \BOthers {.}}{%
{\protect \APACyear {2016}}%
}]{%
chin2016successful}
\APACinsertmetastar {%
chin2016successful}%
\begin{APACrefauthors}%
Chin, S\BPBI O.%
, Keum, C.%
, Woo, J.%
, Park, J.%
, Choi, H\BPBI J.%
, Woo, J\BHBI t.%
\BCBL {}\ \BBA {} Rhee, S\BPBI Y.%
\end{APACrefauthors}%
\unskip\
\newblock
\APACrefYearMonthDay{2016}{}{}.
\newblock
{\BBOQ}\APACrefatitle {Successful weight reduction and maintenance by using a
  smartphone application in those with overweight and obesity} {Successful
  weight reduction and maintenance by using a smartphone application in those
  with overweight and obesity}.{\BBCQ}
\newblock
\APACjournalVolNumPages{Scientific reports}{6}{1}{1--8}.
\PrintBackRefs{\CurrentBib}

\bibitem [\protect \citeauthoryear {%
Ciotti%
\ \protect \BOthers {.}}{%
Ciotti%
\ \protect \BOthers {.}}{%
{\protect \APACyear {2019}}%
}]{%
ciotti2019covid}
\APACinsertmetastar {%
ciotti2019covid}%
\begin{APACrefauthors}%
Ciotti, M.%
, Angeletti, S.%
, Minieri, M.%
, Giovannetti, M.%
, Benvenuto, D.%
, Pascarella, S.%
\BDBL {}Ciccozzi, M.%
\end{APACrefauthors}%
\unskip\
\newblock
\APACrefYearMonthDay{2019}{}{}.
\newblock
{\BBOQ}\APACrefatitle {COVID-19 outbreak: an overview} {Covid-19 outbreak: an
  overview}.{\BBCQ}
\newblock
\APACjournalVolNumPages{Chemotherapy}{64}{5-6}{215--223}.
\PrintBackRefs{\CurrentBib}

\bibitem [\protect \citeauthoryear {%
Colwell%
}{%
Colwell%
}{%
{\protect \APACyear {1996}}%
}]{%
colwell1996global}
\APACinsertmetastar {%
colwell1996global}%
\begin{APACrefauthors}%
Colwell, R\BPBI R.%
\end{APACrefauthors}%
\unskip\
\newblock
\APACrefYearMonthDay{1996}{}{}.
\newblock
{\BBOQ}\APACrefatitle {Global climate and infectious disease: the cholera
  paradigm} {Global climate and infectious disease: the cholera
  paradigm}.{\BBCQ}
\newblock
\APACjournalVolNumPages{Science}{274}{5295}{2025--2031}.
\PrintBackRefs{\CurrentBib}

\bibitem [\protect \citeauthoryear {%
Cunha%
}{%
Cunha%
}{%
{\protect \APACyear {2004}}%
}]{%
cunha2004influenza}
\APACinsertmetastar {%
cunha2004influenza}%
\begin{APACrefauthors}%
Cunha, B\BPBI A.%
\end{APACrefauthors}%
\unskip\
\newblock
\APACrefYearMonthDay{2004}{}{}.
\newblock
{\BBOQ}\APACrefatitle {Influenza: historical aspects of epidemics and
  pandemics} {Influenza: historical aspects of epidemics and pandemics}.{\BBCQ}
\newblock
\APACjournalVolNumPages{Infectious Disease Clinics}{18}{1}{141--155}.
\PrintBackRefs{\CurrentBib}

\bibitem [\protect \citeauthoryear {%
Dong%
, Du%
\BCBL {}\ \BBA {} Gardner%
}{%
Dong%
\ \protect \BOthers {.}}{%
{\protect \APACyear {2020}}%
}]{%
dong2020interactive}
\APACinsertmetastar {%
dong2020interactive}%
\begin{APACrefauthors}%
Dong, E.%
, Du, H.%
\BCBL {}\ \BBA {} Gardner, L.%
\end{APACrefauthors}%
\unskip\
\newblock
\APACrefYearMonthDay{2020}{}{}.
\newblock
{\BBOQ}\APACrefatitle {An interactive web-based dashboard to track COVID-19 in
  real time} {An interactive web-based dashboard to track covid-19 in real
  time}.{\BBCQ}
\newblock
\APACjournalVolNumPages{The Lancet infectious diseases}{}{}{}.
\PrintBackRefs{\CurrentBib}

\bibitem [\protect \citeauthoryear {%
Eastwood%
, Durrheim%
, Butler%
\BCBL {}\ \BBA {} Jones%
}{%
Eastwood%
\ \protect \BOthers {.}}{%
{\protect \APACyear {2010}}%
}]{%
eastwood2010responses}
\APACinsertmetastar {%
eastwood2010responses}%
\begin{APACrefauthors}%
Eastwood, K.%
, Durrheim, D\BPBI N.%
, Butler, M.%
\BCBL {}\ \BBA {} Jones, A.%
\end{APACrefauthors}%
\unskip\
\newblock
\APACrefYearMonthDay{2010}{}{}.
\newblock
{\BBOQ}\APACrefatitle {Responses to pandemic (H1N1) 2009, Australia} {Responses
  to pandemic (h1n1) 2009, australia}.{\BBCQ}
\newblock
\APACjournalVolNumPages{Emerging Infectious Diseases}{16}{8}{1211}.
\PrintBackRefs{\CurrentBib}

\bibitem [\protect \citeauthoryear {%
Forni%
\ \BBA {} Mantovani%
}{%
Forni%
\ \BBA {} Mantovani%
}{%
{\protect \APACyear {2021}}%
}]{%
forni2021covid}
\APACinsertmetastar {%
forni2021covid}%
\begin{APACrefauthors}%
Forni, G.%
\BCBT {}\ \BBA {} Mantovani, A.%
\end{APACrefauthors}%
\unskip\
\newblock
\APACrefYearMonthDay{2021}{}{}.
\newblock
{\BBOQ}\APACrefatitle {COVID-19 vaccines: where we stand and challenges ahead}
  {Covid-19 vaccines: where we stand and challenges ahead}.{\BBCQ}
\newblock
\APACjournalVolNumPages{Cell Death \& Differentiation}{28}{2}{626--639}.
\PrintBackRefs{\CurrentBib}

\bibitem [\protect \citeauthoryear {%
G{\"U}NER%
, Hasano{\u{g}}lu%
\BCBL {}\ \BBA {} Akta{\c{s}}%
}{%
G{\"U}NER%
\ \protect \BOthers {.}}{%
{\protect \APACyear {2020}}%
}]{%
guner2020covid}
\APACinsertmetastar {%
guner2020covid}%
\begin{APACrefauthors}%
G{\"U}NER, H\BPBI R.%
, Hasano{\u{g}}lu, I.%
\BCBL {}\ \BBA {} Akta{\c{s}}, F.%
\end{APACrefauthors}%
\unskip\
\newblock
\APACrefYearMonthDay{2020}{}{}.
\newblock
{\BBOQ}\APACrefatitle {COVID-19: Prevention and control measures in community}
  {Covid-19: Prevention and control measures in community}.{\BBCQ}
\newblock
\APACjournalVolNumPages{Turkish Journal of medical
  sciences}{50}{SI-1}{571--577}.
\PrintBackRefs{\CurrentBib}

\bibitem [\protect \citeauthoryear {%
Hao%
\ \protect \BOthers {.}}{%
Hao%
\ \protect \BOthers {.}}{%
{\protect \APACyear {2020}}%
}]{%
hao2020psychiatric}
\APACinsertmetastar {%
hao2020psychiatric}%
\begin{APACrefauthors}%
Hao, F.%
, Tan, W.%
, Jiang, L.%
, Zhang, L.%
, Zhao, X.%
, Zou, Y.%
\BDBL {}others%
\end{APACrefauthors}%
\unskip\
\newblock
\APACrefYearMonthDay{2020}{}{}.
\newblock
{\BBOQ}\APACrefatitle {Do psychiatric patients experience more psychiatric
  symptoms during COVID-19 pandemic and lockdown? A case-control study with
  service and research implications for immunopsychiatry} {Do psychiatric
  patients experience more psychiatric symptoms during covid-19 pandemic and
  lockdown? a case-control study with service and research implications for
  immunopsychiatry}.{\BBCQ}
\newblock
\APACjournalVolNumPages{Brain, behavior, and immunity}{87}{}{100--106}.
\PrintBackRefs{\CurrentBib}

\bibitem [\protect \citeauthoryear {%
Hays%
}{%
Hays%
}{%
{\protect \APACyear {2005}}%
}]{%
hays2005epidemics}
\APACinsertmetastar {%
hays2005epidemics}%
\begin{APACrefauthors}%
Hays, J\BPBI N.%
\end{APACrefauthors}%
\unskip\
\newblock
\APACrefYear{2005}.
\newblock
\APACrefbtitle {Epidemics and pandemics: their impacts on human history}
  {Epidemics and pandemics: their impacts on human history}.
\newblock
\APACaddressPublisher{}{Abc-clio}.
\PrintBackRefs{\CurrentBib}

\bibitem [\protect \citeauthoryear {%
Kong%
\ \BBA {} Tan%
}{%
Kong%
\ \BBA {} Tan%
}{%
{\protect \APACyear {2012}}%
}]{%
kong2012dietcam}
\APACinsertmetastar {%
kong2012dietcam}%
\begin{APACrefauthors}%
Kong, F.%
\BCBT {}\ \BBA {} Tan, J.%
\end{APACrefauthors}%
\unskip\
\newblock
\APACrefYearMonthDay{2012}{}{}.
\newblock
{\BBOQ}\APACrefatitle {DietCam: Automatic dietary assessment with mobile camera
  phones} {Dietcam: Automatic dietary assessment with mobile camera
  phones}.{\BBCQ}
\newblock
\APACjournalVolNumPages{Pervasive and Mobile Computing}{8}{1}{147--163}.
\PrintBackRefs{\CurrentBib}

\bibitem [\protect \citeauthoryear {%
Lacitignola%
\ \BBA {} Saccomandi%
}{%
Lacitignola%
\ \BBA {} Saccomandi%
}{%
{\protect \APACyear {2021}}%
}]{%
lacitignola2021managing}
\APACinsertmetastar {%
lacitignola2021managing}%
\begin{APACrefauthors}%
Lacitignola, D.%
\BCBT {}\ \BBA {} Saccomandi, G.%
\end{APACrefauthors}%
\unskip\
\newblock
\APACrefYearMonthDay{2021}{}{}.
\newblock
{\BBOQ}\APACrefatitle {Managing awareness can avoid hysteresis in disease
  spread: an application to coronavirus Covid-19} {Managing awareness can avoid
  hysteresis in disease spread: an application to coronavirus covid-19}.{\BBCQ}
\newblock
\APACjournalVolNumPages{Chaos, Solitons \& Fractals}{144}{}{110739}.
\PrintBackRefs{\CurrentBib}

\bibitem [\protect \citeauthoryear {%
Li%
\ \protect \BOthers {.}}{%
Li%
\ \protect \BOthers {.}}{%
{\protect \APACyear {2007}}%
}]{%
li2007origin}
\APACinsertmetastar {%
li2007origin}%
\begin{APACrefauthors}%
Li, Y.%
, Carroll, D\BPBI S.%
, Gardner, S\BPBI N.%
, Walsh, M\BPBI C.%
, Vitalis, E\BPBI A.%
\BCBL {}\ \BBA {} Damon, I\BPBI K.%
\end{APACrefauthors}%
\unskip\
\newblock
\APACrefYearMonthDay{2007}{}{}.
\newblock
{\BBOQ}\APACrefatitle {On the origin of smallpox: correlating variola
  phylogenics with historical smallpox records} {On the origin of smallpox:
  correlating variola phylogenics with historical smallpox records}.{\BBCQ}
\newblock
\APACjournalVolNumPages{Proceedings of the National Academy of
  Sciences}{104}{40}{15787--15792}.
\PrintBackRefs{\CurrentBib}

\bibitem [\protect \citeauthoryear {%
Manuel%
, Jos{\'e}%
, Manuel%
, Iv{\'a}n%
\BCBL {}\ \BBA {} Baltasar%
}{%
Manuel%
\ \protect \BOthers {.}}{%
{\protect \APACyear {2019}}%
}]{%
manuel2019simplifying}
\APACinsertmetastar {%
manuel2019simplifying}%
\begin{APACrefauthors}%
Manuel, P\BHBI C\BPBI V.%
, Jos{\'e}, P\BHBI C\BPBI I.%
, Manuel, F\BHBI M.%
, Iv{\'a}n, M\BHBI O.%
\BCBL {}\ \BBA {} Baltasar, F\BHBI M.%
\end{APACrefauthors}%
\unskip\
\newblock
\APACrefYearMonthDay{2019}{}{}.
\newblock
{\BBOQ}\APACrefatitle {Simplifying the Creation of Adventure Serious Games with
  Educational-Oriented Features} {Simplifying the creation of adventure serious
  games with educational-oriented features}.{\BBCQ}
\newblock
\APACjournalVolNumPages{Journal of Educational Technology \&
  Society}{22}{3}{32--46}.
\PrintBackRefs{\CurrentBib}

\bibitem [\protect \citeauthoryear {%
Matsumura%
\ \BBA {} Yamakoshi%
}{%
Matsumura%
\ \BBA {} Yamakoshi%
}{%
{\protect \APACyear {2013}}%
}]{%
matsumura2013iphysiometer}
\APACinsertmetastar {%
matsumura2013iphysiometer}%
\begin{APACrefauthors}%
Matsumura, K.%
\BCBT {}\ \BBA {} Yamakoshi, T.%
\end{APACrefauthors}%
\unskip\
\newblock
\APACrefYearMonthDay{2013}{}{}.
\newblock
{\BBOQ}\APACrefatitle {iPhysioMeter: a new approach for measuring heart rate
  and normalized pulse volume using only a smartphone} {iphysiometer: a new
  approach for measuring heart rate and normalized pulse volume using only a
  smartphone}.{\BBCQ}
\newblock
\APACjournalVolNumPages{Behavior research methods}{45}{4}{1272--1278}.
\PrintBackRefs{\CurrentBib}

\bibitem [\protect \citeauthoryear {%
Mutreja%
\ \protect \BOthers {.}}{%
Mutreja%
\ \protect \BOthers {.}}{%
{\protect \APACyear {2011}}%
}]{%
mutreja2011evidence}
\APACinsertmetastar {%
mutreja2011evidence}%
\begin{APACrefauthors}%
Mutreja, A.%
, Kim, D\BPBI W.%
, Thomson, N\BPBI R.%
, Connor, T\BPBI R.%
, Lee, J\BPBI H.%
, Kariuki, S.%
\BDBL {}others%
\end{APACrefauthors}%
\unskip\
\newblock
\APACrefYearMonthDay{2011}{}{}.
\newblock
{\BBOQ}\APACrefatitle {Evidence for several waves of global transmission in the
  seventh cholera pandemic} {Evidence for several waves of global transmission
  in the seventh cholera pandemic}.{\BBCQ}
\newblock
\APACjournalVolNumPages{Nature}{477}{7365}{462--465}.
\PrintBackRefs{\CurrentBib}

\bibitem [\protect \citeauthoryear {%
Mytton%
, Rutter%
\BCBL {}\ \BBA {} Donaldson%
}{%
Mytton%
\ \protect \BOthers {.}}{%
{\protect \APACyear {2012}}%
}]{%
mytton2012influenza}
\APACinsertmetastar {%
mytton2012influenza}%
\begin{APACrefauthors}%
Mytton, O.%
, Rutter, P.%
\BCBL {}\ \BBA {} Donaldson, L.%
\end{APACrefauthors}%
\unskip\
\newblock
\APACrefYearMonthDay{2012}{}{}.
\newblock
{\BBOQ}\APACrefatitle {Influenza A (H1N1) pdm09 in England, 2009 to 2011: a
  greater burden of severe illness in the year after the pandemic than in the
  pandemic year} {Influenza a (h1n1) pdm09 in england, 2009 to 2011: a greater
  burden of severe illness in the year after the pandemic than in the pandemic
  year}.{\BBCQ}
\newblock
\APACjournalVolNumPages{Euro surveillance}{17}{14}{11--19}.
\PrintBackRefs{\CurrentBib}

\bibitem [\protect \citeauthoryear {%
Nussbaumer-Streit%
\ \protect \BOthers {.}}{%
Nussbaumer-Streit%
\ \protect \BOthers {.}}{%
{\protect \APACyear {2020}}%
}]{%
nussbaumer2020quarantine}
\APACinsertmetastar {%
nussbaumer2020quarantine}%
\begin{APACrefauthors}%
Nussbaumer-Streit, B.%
, Mayr, V.%
, Dobrescu, A\BPBI I.%
, Chapman, A.%
, Persad, E.%
, Klerings, I.%
\BDBL {}others%
\end{APACrefauthors}%
\unskip\
\newblock
\APACrefYearMonthDay{2020}{}{}.
\newblock
{\BBOQ}\APACrefatitle {Quarantine alone or in combination with other public
  health measures to control COVID-19: a rapid review} {Quarantine alone or in
  combination with other public health measures to control covid-19: a rapid
  review}.{\BBCQ}
\newblock
\APACjournalVolNumPages{Cochrane Database of Systematic Reviews}{}{9}{}.
\PrintBackRefs{\CurrentBib}

\bibitem [\protect \citeauthoryear {%
Orji%
, Vassileva%
\BCBL {}\ \BBA {} Mandryk%
}{%
Orji%
\ \protect \BOthers {.}}{%
{\protect \APACyear {2013}}%
}]{%
orji2013lunchtime}
\APACinsertmetastar {%
orji2013lunchtime}%
\begin{APACrefauthors}%
Orji, R.%
, Vassileva, J.%
\BCBL {}\ \BBA {} Mandryk, R\BPBI L.%
\end{APACrefauthors}%
\unskip\
\newblock
\APACrefYearMonthDay{2013}{}{}.
\newblock
{\BBOQ}\APACrefatitle {LunchTime: a slow-casual game for long-term dietary
  behavior change} {Lunchtime: a slow-casual game for long-term dietary
  behavior change}.{\BBCQ}
\newblock
\APACjournalVolNumPages{Personal and Ubiquitous Computing}{17}{6}{1211--1221}.
\PrintBackRefs{\CurrentBib}

\bibitem [\protect \citeauthoryear {%
Park%
, Koo%
, Cho%
\BCBL {}\ \BBA {} Bae%
}{%
Park%
\ \protect \BOthers {.}}{%
{\protect \APACyear {2015}}%
}]{%
park2015snackbreaker}
\APACinsertmetastar {%
park2015snackbreaker}%
\begin{APACrefauthors}%
Park, J.%
, Koo, B\BHBI c.%
, Cho, J.%
\BCBL {}\ \BBA {} Bae, B\BHBI C.%
\end{APACrefauthors}%
\unskip\
\newblock
\APACrefYearMonthDay{2015}{}{}.
\newblock
{\BBOQ}\APACrefatitle {SnackBreaker: A Game Promoting Healthy Choice of Snack
  Foods} {Snackbreaker: A game promoting healthy choice of snack foods}.{\BBCQ}
\newblock
\BIn{} \APACrefbtitle {Proceedings of the 2015 Annual Symposium on
  Computer-Human Interaction in Play} {Proceedings of the 2015 annual symposium
  on computer-human interaction in play}\ (\BPGS\ 673--678).
\PrintBackRefs{\CurrentBib}

\bibitem [\protect \citeauthoryear {%
Richardson%
\ \protect \BOthers {.}}{%
Richardson%
\ \protect \BOthers {.}}{%
{\protect \APACyear {2016}}%
}]{%
richardson2016biosocial}
\APACinsertmetastar {%
richardson2016biosocial}%
\begin{APACrefauthors}%
Richardson, E\BPBI T.%
, Barrie, M\BPBI B.%
, Kelly, J\BPBI D.%
, Dibba, Y.%
, Koedoyoma, S.%
\BCBL {}\ \BBA {} Farmer, P\BPBI E.%
\end{APACrefauthors}%
\unskip\
\newblock
\APACrefYearMonthDay{2016}{}{}.
\newblock
{\BBOQ}\APACrefatitle {Biosocial approaches to the 2013-2016 Ebola pandemic}
  {Biosocial approaches to the 2013-2016 ebola pandemic}.{\BBCQ}
\newblock
\APACjournalVolNumPages{Health and human rights}{18}{1}{115}.
\PrintBackRefs{\CurrentBib}

\bibitem [\protect \citeauthoryear {%
Rozhnova%
\ \protect \BOthers {.}}{%
Rozhnova%
\ \protect \BOthers {.}}{%
{\protect \APACyear {2021}}%
}]{%
rozhnova2021model}
\APACinsertmetastar {%
rozhnova2021model}%
\begin{APACrefauthors}%
Rozhnova, G.%
, van Dorp, C\BPBI H.%
, Bruijning-Verhagen, P.%
, Bootsma, M\BPBI C.%
, van~de Wijgert, J\BPBI H.%
, Bonten, M\BPBI J.%
\BCBL {}\ \BBA {} Kretzschmar, M\BPBI E.%
\end{APACrefauthors}%
\unskip\
\newblock
\APACrefYearMonthDay{2021}{}{}.
\newblock
{\BBOQ}\APACrefatitle {Model-based evaluation of school-and non-school-related
  measures to control the COVID-19 pandemic} {Model-based evaluation of
  school-and non-school-related measures to control the covid-19
  pandemic}.{\BBCQ}
\newblock
\APACjournalVolNumPages{Nature communications}{12}{1}{1--11}.
\PrintBackRefs{\CurrentBib}

\bibitem [\protect \citeauthoryear {%
Santos%
, Carvalho%
, Costa%
, Viana%
\BCBL {}\ \BBA {} Rivero%
}{%
Santos%
\ \protect \BOthers {.}}{%
{\protect \APACyear {2019}}%
}]{%
santos2019risking}
\APACinsertmetastar {%
santos2019risking}%
\begin{APACrefauthors}%
Santos, S.%
, Carvalho, F.%
, Costa, Y.%
, Viana, D.%
\BCBL {}\ \BBA {} Rivero, L.%
\end{APACrefauthors}%
\unskip\
\newblock
\APACrefYearMonthDay{2019}{}{}.
\newblock
{\BBOQ}\APACrefatitle {Risking: A game for teaching risk management in software
  projects} {Risking: A game for teaching risk management in software
  projects}.{\BBCQ}
\newblock
\BIn{} \APACrefbtitle {Proceedings of the XVIII Brazilian Symposium on Software
  Quality} {Proceedings of the xviii brazilian symposium on software quality}\
  (\BPGS\ 188--197).
\PrintBackRefs{\CurrentBib}

\bibitem [\protect \citeauthoryear {%
Sepulveda%
, Valdespino%
\BCBL {}\ \BBA {} Garcia-Garcia%
}{%
Sepulveda%
\ \protect \BOthers {.}}{%
{\protect \APACyear {2006}}%
}]{%
sepulveda2006cholera}
\APACinsertmetastar {%
sepulveda2006cholera}%
\begin{APACrefauthors}%
Sepulveda, J.%
, Valdespino, J\BPBI L.%
\BCBL {}\ \BBA {} Garcia-Garcia, L.%
\end{APACrefauthors}%
\unskip\
\newblock
\APACrefYearMonthDay{2006}{}{}.
\newblock
{\BBOQ}\APACrefatitle {Cholera in Mexico: the paradoxical benefits of the last
  pandemic} {Cholera in mexico: the paradoxical benefits of the last
  pandemic}.{\BBCQ}
\newblock
\APACjournalVolNumPages{International Journal of Infectious
  Diseases}{10}{1}{4--13}.
\PrintBackRefs{\CurrentBib}

\bibitem [\protect \citeauthoryear {%
Sindi%
\ \protect \BOthers {.}}{%
Sindi%
\ \protect \BOthers {.}}{%
{\protect \APACyear {2015}}%
}]{%
sindi2015caide}
\APACinsertmetastar {%
sindi2015caide}%
\begin{APACrefauthors}%
Sindi, S.%
, Calov, E.%
, Fokkens, J.%
, Ngandu, T.%
, Soininen, H.%
, Tuomilehto, J.%
\BCBL {}\ \BBA {} Kivipelto, M.%
\end{APACrefauthors}%
\unskip\
\newblock
\APACrefYearMonthDay{2015}{}{}.
\newblock
{\BBOQ}\APACrefatitle {The CAIDE Dementia Risk Score App: The development of an
  evidence-based mobile application to predict the risk of dementia} {The caide
  dementia risk score app: The development of an evidence-based mobile
  application to predict the risk of dementia}.{\BBCQ}
\newblock
\APACjournalVolNumPages{Alzheimer's \& Dementia: Diagnosis, Assessment \&
  Disease Monitoring}{1}{3}{328--333}.
\PrintBackRefs{\CurrentBib}

\bibitem [\protect \citeauthoryear {%
Sohrabi%
\ \protect \BOthers {.}}{%
Sohrabi%
\ \protect \BOthers {.}}{%
{\protect \APACyear {2020}}%
}]{%
sohrabi2020world}
\APACinsertmetastar {%
sohrabi2020world}%
\begin{APACrefauthors}%
Sohrabi, C.%
, Alsafi, Z.%
, O’Neill, N.%
, Khan, M.%
, Kerwan, A.%
, Al-Jabir, A.%
\BDBL {}Agha, R.%
\end{APACrefauthors}%
\unskip\
\newblock
\APACrefYearMonthDay{2020}{}{}.
\newblock
{\BBOQ}\APACrefatitle {World Health Organization declares global emergency: A
  review of the 2019 novel coronavirus (COVID-19)} {World health organization
  declares global emergency: A review of the 2019 novel coronavirus
  (covid-19)}.{\BBCQ}
\newblock
\APACjournalVolNumPages{International Journal of Surgery}{}{}{}.
\PrintBackRefs{\CurrentBib}

\bibitem [\protect \citeauthoryear {%
Valenza%
, da Silva~Hounsell%
\BCBL {}\ \BBA {} Gasparini%
}{%
Valenza%
\ \protect \BOthers {.}}{%
{\protect \APACyear {2019}}%
}]{%
valenza2019serious}
\APACinsertmetastar {%
valenza2019serious}%
\begin{APACrefauthors}%
Valenza, M\BPBI V.%
, da Silva~Hounsell, M.%
\BCBL {}\ \BBA {} Gasparini, I.%
\end{APACrefauthors}%
\unskip\
\newblock
\APACrefYearMonthDay{2019}{}{}.
\newblock
\APACrefbtitle {Serious game design for children: Validating a set of
  guidelines} {Serious game design for children: Validating a set of
  guidelines}\ (\BVOL~22)\ (\BNUM~3).
\newblock
\APACaddressPublisher{}{JSTOR}.
\PrintBackRefs{\CurrentBib}

\bibitem [\protect \citeauthoryear {%
Venigalla%
, Chimalakonda%
\BCBL {}\ \BBA {} Vagavolu%
}{%
Venigalla%
\ \protect \BOthers {.}}{%
{\protect \APACyear {2020}}%
}]{%
venigalla2020mood}
\APACinsertmetastar {%
venigalla2020mood}%
\begin{APACrefauthors}%
Venigalla, A\BPBI S\BPBI M.%
, Chimalakonda, S.%
\BCBL {}\ \BBA {} Vagavolu, D.%
\end{APACrefauthors}%
\unskip\
\newblock
\APACrefYearMonthDay{2020}{}{}.
\newblock
{\BBOQ}\APACrefatitle {Mood of India During Covid-19-An Interactive Web Portal
  Based on Emotion Analysis of Twitter Data} {Mood of india during covid-19-an
  interactive web portal based on emotion analysis of twitter data}.{\BBCQ}
\newblock
\BIn{} \APACrefbtitle {Conference Companion Publication of the 2020 on Computer
  Supported Cooperative Work and Social Computing} {Conference companion
  publication of the 2020 on computer supported cooperative work and social
  computing}\ (\BPGS\ 65--68).
\PrintBackRefs{\CurrentBib}

\bibitem [\protect \citeauthoryear {%
H\BHBI Y.~Wang%
\ \protect \BOthers {.}}{%
H\BHBI Y.~Wang%
\ \protect \BOthers {.}}{%
{\protect \APACyear {2020}}%
}]{%
wang2020potential}
\APACinsertmetastar {%
wang2020potential}%
\begin{APACrefauthors}%
Wang, H\BHBI Y.%
, Li, X\BHBI L.%
, Yan, Z\BHBI R.%
, Sun, X\BHBI P.%
, Han, J.%
\BCBL {}\ \BBA {} Zhang, B\BHBI W.%
\end{APACrefauthors}%
\unskip\
\newblock
\APACrefYearMonthDay{2020}{}{}.
\newblock
{\BBOQ}\APACrefatitle {Potential neurological symptoms of COVID-19} {Potential
  neurological symptoms of covid-19}.{\BBCQ}
\newblock
\APACjournalVolNumPages{Therapeutic advances in neurological
  disorders}{13}{}{1756286420917830}.
\PrintBackRefs{\CurrentBib}

\bibitem [\protect \citeauthoryear {%
Y\BHBI H.~Wang%
}{%
Y\BHBI H.~Wang%
}{%
{\protect \APACyear {2020}}%
}]{%
wang2020integrating}
\APACinsertmetastar {%
wang2020integrating}%
\begin{APACrefauthors}%
Wang, Y\BHBI H.%
\end{APACrefauthors}%
\unskip\
\newblock
\APACrefYearMonthDay{2020}{}{}.
\newblock
{\BBOQ}\APACrefatitle {Integrating Games, e-Books and AR Techniques to Support
  Project-based Science Learning} {Integrating games, e-books and ar techniques
  to support project-based science learning}.{\BBCQ}
\newblock
\APACjournalVolNumPages{Educational Technology \& Society}{23}{3}{53--67}.
\PrintBackRefs{\CurrentBib}

\bibitem [\protect \citeauthoryear {%
Westera%
}{%
Westera%
}{%
{\protect \APACyear {2019}}%
}]{%
westera2019and}
\APACinsertmetastar {%
westera2019and}%
\begin{APACrefauthors}%
Westera, W.%
\end{APACrefauthors}%
\unskip\
\newblock
\APACrefYearMonthDay{2019}{}{}.
\newblock
{\BBOQ}\APACrefatitle {Why and how serious games can become far more effective:
  Accommodating productive learning experiences, learner motivation and the
  monitoring of learning gains} {Why and how serious games can become far more
  effective: Accommodating productive learning experiences, learner motivation
  and the monitoring of learning gains}.{\BBCQ}
\newblock
\APACjournalVolNumPages{Journal of Educational Technology \&
  Society}{22}{1}{59--69}.
\PrintBackRefs{\CurrentBib}

\bibitem [\protect \citeauthoryear {%
Xu%
\ \protect \BOthers {.}}{%
Xu%
\ \protect \BOthers {.}}{%
{\protect \APACyear {2010}}%
}]{%
xu2010structural}
\APACinsertmetastar {%
xu2010structural}%
\begin{APACrefauthors}%
Xu, R.%
, Ekiert, D\BPBI C.%
, Krause, J\BPBI C.%
, Hai, R.%
, Crowe, J\BPBI E.%
\BCBL {}\ \BBA {} Wilson, I\BPBI A.%
\end{APACrefauthors}%
\unskip\
\newblock
\APACrefYearMonthDay{2010}{}{}.
\newblock
{\BBOQ}\APACrefatitle {Structural basis of preexisting immunity to the 2009
  H1N1 pandemic influenza virus} {Structural basis of preexisting immunity to
  the 2009 h1n1 pandemic influenza virus}.{\BBCQ}
\newblock
\APACjournalVolNumPages{Science}{328}{5976}{357--360}.
\PrintBackRefs{\CurrentBib}

\bibitem [\protect \citeauthoryear {%
Yuki%
, Fujiogi%
\BCBL {}\ \BBA {} Koutsogiannaki%
}{%
Yuki%
\ \protect \BOthers {.}}{%
{\protect \APACyear {2020}}%
}]{%
yuki2020covid}
\APACinsertmetastar {%
yuki2020covid}%
\begin{APACrefauthors}%
Yuki, K.%
, Fujiogi, M.%
\BCBL {}\ \BBA {} Koutsogiannaki, S.%
\end{APACrefauthors}%
\unskip\
\newblock
\APACrefYearMonthDay{2020}{}{}.
\newblock
{\BBOQ}\APACrefatitle {COVID-19 pathophysiology: A review} {Covid-19
  pathophysiology: A review}.{\BBCQ}
\newblock
\APACjournalVolNumPages{Clinical immunology}{}{}{108427}.
\PrintBackRefs{\CurrentBib}

\bibitem [\protect \citeauthoryear {%
Zheng%
, Ma%
, Zhang%
\BCBL {}\ \BBA {} Xie%
}{%
Zheng%
\ \protect \BOthers {.}}{%
{\protect \APACyear {2020}}%
}]{%
zheng2020covid}
\APACinsertmetastar {%
zheng2020covid}%
\begin{APACrefauthors}%
Zheng, Y\BHBI Y.%
, Ma, Y\BHBI T.%
, Zhang, J\BHBI Y.%
\BCBL {}\ \BBA {} Xie, X.%
\end{APACrefauthors}%
\unskip\
\newblock
\APACrefYearMonthDay{2020}{}{}.
\newblock
{\BBOQ}\APACrefatitle {COVID-19 and the cardiovascular system} {Covid-19 and
  the cardiovascular system}.{\BBCQ}
\newblock
\APACjournalVolNumPages{Nature Reviews Cardiology}{}{}{1--2}.
\PrintBackRefs{\CurrentBib}

\bibitem [\protect \citeauthoryear {%
Zhou%
\ \protect \BOthers {.}}{%
Zhou%
\ \protect \BOthers {.}}{%
{\protect \APACyear {2020}}%
}]{%
zhou2020clinical}
\APACinsertmetastar {%
zhou2020clinical}%
\begin{APACrefauthors}%
Zhou, F.%
, Yu, T.%
, Du, R.%
, Fan, G.%
, Liu, Y.%
, Liu, Z.%
\BDBL {}others%
\end{APACrefauthors}%
\unskip\
\newblock
\APACrefYearMonthDay{2020}{}{}.
\newblock
{\BBOQ}\APACrefatitle {Clinical course and risk factors for mortality of adult
  inpatients with COVID-19 in Wuhan, China: a retrospective cohort study}
  {Clinical course and risk factors for mortality of adult inpatients with
  covid-19 in wuhan, china: a retrospective cohort study}.{\BBCQ}
\newblock
\APACjournalVolNumPages{The Lancet}{}{}{}.
\PrintBackRefs{\CurrentBib}

\end{thebibliography}

\end{document}